\documentclass[pra,twocolumn,amsmath,amssymb,superscriptaddress,groupedaddress,showpacs]{revtex4}

\usepackage[letterpaper]{geometry}

\usepackage{amsmath,array,amssymb}
\usepackage{graphicx,color}

\usepackage{braket,paralist}
\usepackage{indentfirst}
 \usepackage{epsfig}

\newcommand{\sa}{\psi_A}
\newcommand{\sbp}{\psi_B^{(p)}}
\newcommand{\sbn}{\psi_B^{(n)}}
\renewcommand{\vec}[1]{\boldsymbol{#1}}
\newcommand{\figref}[1]{Fig.~\ref{#1}}
\newcommand{\etal}{\textit{ et.~al.~}}

\begin{document}
\title{Playing the Aharon-Vaidman quantum game with a Young type photonic qutrit}
\author{Piotr Kolenderski}
\email{kolenderski@fizyka.umk.pl}
\affiliation{Institute for Quantum Computing, University of Waterloo, 200 University Ave. West, Waterloo, Ontario, CA
N2L 3G1}
\affiliation{Institute of Physics, Nicolaus Copernicus University, Grudzi\c{a}dzka 5, 87-100 Toru{\'n}, Poland}
\author{Urbasi Sinha}
\affiliation{Institute for Quantum Computing, University of Waterloo, 200 University Ave. West, Waterloo, Ontario, CA
N2L 3G1}
\affiliation{Raman Research Institute, Sadashivanagar, Bangalore 560080, India}
\author{Li Youning}
\affiliation{Department of Physics, Tsinghua University, Beijing 100084, P. R. China}
\author{Tong Zhao}
\author{Matthew Volpini}
\affiliation{Institute for Quantum Computing, University of Waterloo, 200 University Ave. West, Waterloo, Ontario, CA
N2L 3G1}
\author{Ad\'an Cabello}
 \affiliation{Departamento de F\'{\i}sica Aplicada II, Universidad de Sevilla, E-41012 Sevilla, Spain}
 \affiliation{Department of Physics, Stockholm University, S-10691 Stockholm, Sweden}
\author{Raymond Laflamme}
\author{Thomas Jennewein} 
 \affiliation{Institute for Quantum Computing, University of Waterloo, 200 University Ave. West, Waterloo, Ontario, CA
N2L 3G1}


\begin{abstract}
The Aharon-Vaidman (AV) game exemplifies the advantage of using simple quantum systems to outperform classical strategies. We present an experimental test of this advantage by using
a three-state quantum system (qutrit) encoded in a spatial mode of a single photon passing through three slits. The preparation of a particular state is controlled as the photon propagates through the slits by varying the number of open slits and their respective phases. The measurements are achieved by placing detectors in the specific positions in the near and far fields after the slits. This set of tools allowed us to perform tomographic reconstructions of generalized qutrit states, and to implement the quantum version of the AV game with compelling evidence of the quantum advantage.
\end{abstract}

\maketitle

\section{Introduction}

The Aharon-Vaidman game \cite{Aharon2008} is a conceptually simple example of how quantum mechanics can be both beneficial and counter-intuitive. In the classical analog, Alice puts a particle in one of three boxes such that when Bob, who in next turn is allowed to check only two of them, is most likely to find it. Alice who can either accept or not accept a particular game trial wins whenever she accepts a trial in which Bob has also discovered the particle.  Hence, it is obvious that Alice will not use the box that Bob does not have access to and therefore her chance to win is $50\%$. However, the result of the game can be totally different when  the particle is described by laws of quantum mechanics and Alice prepares it in equal superposition of being in each of the boxes. Then her chance to win can reach $100\%$ if she takes a specific projective measurement after Bob's turn.

In this article we present an experimental realization of the AV quantum game using a single photon as the incident particle and a system of three slits in lieu of the boxes. The original three box paradox was proposed by Aharonov and Vaidman in Ref.~\cite{Aharonov1991}. The quantum game \cite{Aharon2008} was conceived much later and exhibits a clear quantum advantage if the game rules are followed with care.  The setup comprising of a single photon source (heralded parametric down conversion source or attenuated laser), triple slit \cite{Sinha2010} and single photon detectors allowed us to perform optimized quantum tomography to characterize the qutrit states and  to play the game correctly in the next step. Some quantum communication protocols that can be considered as quantum games such ascoin tossing \cite{Molina-Terriza2005} and the Byzantine Agreement \cite{Gaertner2008, Gao2008, Gaertner2008a} have been demonstrated previously.  However, the AV game is a specific example of a quantum game where one can demonstrate the quantum advantage playing the game with a single particle at a time in contrast to the entanglement based games  \cite{Molina-Terriza2005, Gaertner2008}. 

\section{Qutrit encoding}

We will begin with the introduction of the concept of the qutrit encoded in spatial degrees of freedom of a single photon \cite{Neves2004, Taguchi2008, Taguchi2009}. In the next step we discuss the AV quantum game  \cite{Aharon2008} and its experimental implementation. Presented implementation of the game is conceptually similar to the Young experiment, where multilevel quantum systems can be encoded in paths related to a photon passing through the slits. Recently this type of quantum state encoding has drawn much interest. In particular, Taguchi \etal used  parametric down conversion source to prepare two qubit \cite{Taguchi2008} and two qutrit  \cite{Taguchi2009} states.  In turn, Lima \etal in Ref.~\cite{Lima2011} demonstrated the seven and eight dimensional state encoding.  Alternative approaches resort to state encoding in various hybrid ways such as energy-time \cite{Thew2004a} or polarization-orbital angular momentum \cite{Nagali2010}. Those implementations were useful to perform a Bell test for energy-time entangled qutrits \cite{Thew2004a} and to demonstrate the optimal cloning strategy \cite{Nagali2010a}, respectively. More recently, the noncontextuality of quantum mechanics was tested based on similar scheme \cite{Lapkiewicz2011, Ahrens2011}.

\begin{figure}
\centering
 \includegraphics[width=\columnwidth]{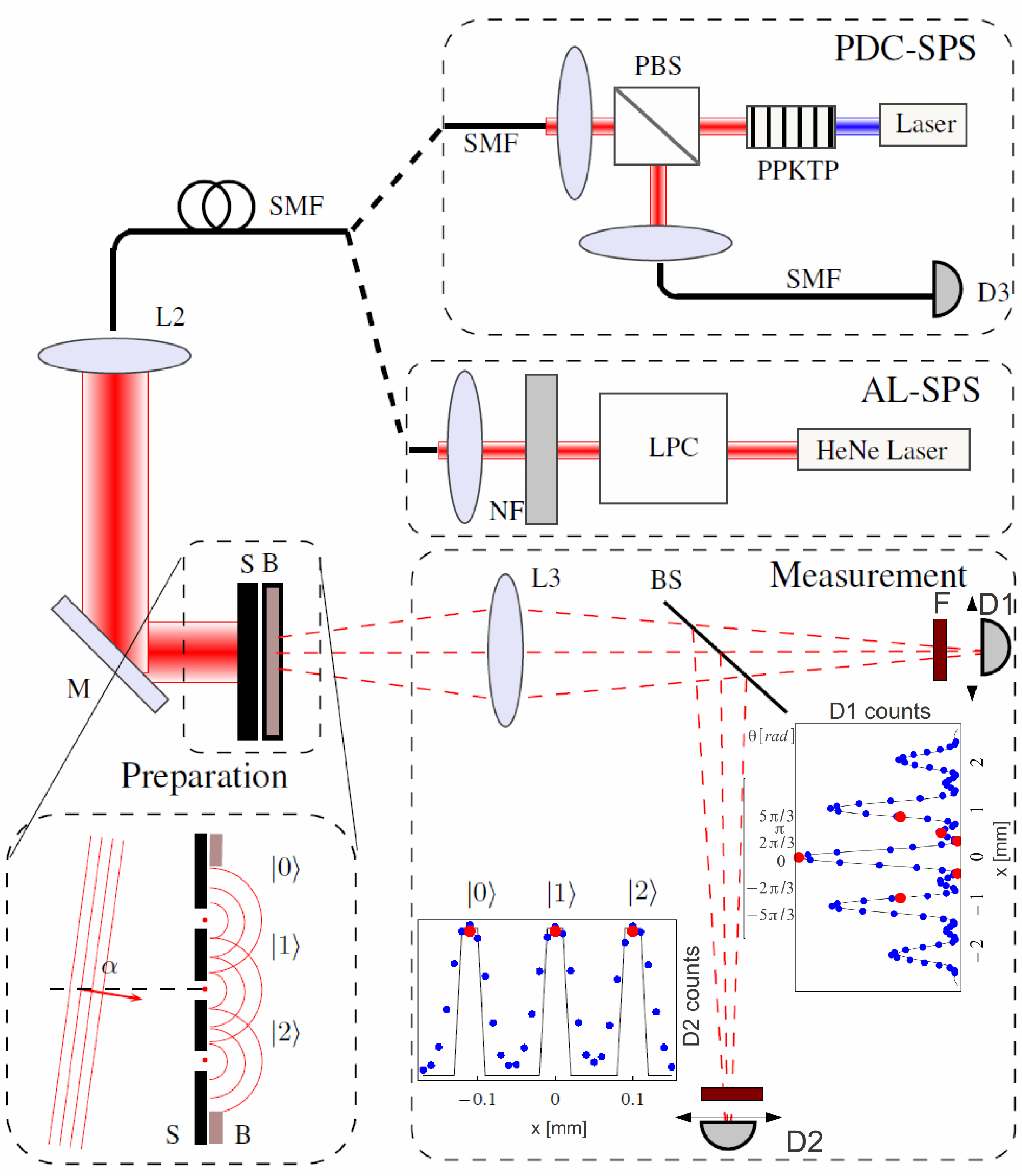}
	\caption{(Color online) Experimental setup. The AL-SPS comprises of HeNe laser ($\lambda=632$~nm), laser power controller (LPC) and neutral filter (NF). The PDC-SPS ($\lambda=810$~nm) is based on PPKTP crystal pumped by blue continuous wave laser. Heralding photon is detected by detector D3. The single photons from both sources are coupled to single mode fibers (SMF). A qutrit is prepared using the blocking mask and three slits. Next the measurement part of the setup comprises of a 2 inch diameter  $f=150$mm lens (L3), 2 inch diameter pellicle beamsplitter (BS), color filters (F) and two detection systems for far (D1) and near filed (D2), each comprised of multimode fiber mounted on a precise motorized stage (Thorlabs ZST13) and a Perkin Elemer avalanche photodiode.}
	\label{fig:experiment}
\end{figure}

The Young type qutrit is realized using triple slits and a single photon source(SPS). The photon's initial spatial mode is  Gaussian with the characteristic diameter much larger than the size of the slits and with the peak intensity coincident with the slit area, see inset in \figref{fig:experiment}. Under these conditions one can consider the state of a photon of wavelength $\lambda$ in the position $\vec{r}=(x,z)$ to be a plane wave $\exp\left(i \vec{k} \vec{r}\right)$ propagating in direction given by  the wave vector $\vec{k}=(k_x,k_z)$ of length $k=2 \pi / \lambda$. Moreover the photon's initial propagation direction is assumed to be paraxial and the distance between the slits larger than their characteristics widths. This allows us to approximate the phase to be constant at each of the slit. 
Hence, the spatial wave function of the photon passing through the $n$th slit  can be written as $\ket{n}=\int_{-\infty}^{\infty}\text{d}x\, S_n(x) \exp\left(i \vec{k} \vec{r}\right)\ket{x}$, where $n=0,1,2$, and $S_n(x)$ stands for the transmission probability amplitude, which we assume is constant on the slit and $0$ elsewhere. 
This means that the total wave function of the photon passing through three slits comprises of three orthogonal contributions. Each of them can be written in momentum representation as \cite{Taguchi2009}:
	$|n\rangle=\int \textrm{d}k_x \tilde{S}_n(x)|k_x \rangle$,
where $\tilde{S}_n(x)= {\sqrt{\frac{a}{2\pi}}}\textrm{sinc}\left(\frac{k_x a}{2}\right)e^{-i n k_x  d}$, $a$ is the slit width and $d$ is the distance between the slits.  These definitions allow us to write the state of the transmitted photon as:
	$\ket{\psi}=\frac{1}{\sqrt{3}}\left(s_1\ket{0}+s_2 \ket{1}+s_3 \ket{2}\right),$
which accounts for the basic definition of a Young type qutrit. Here amplitudes $s_1$, $s_2$ and $s_3$ depend on the transmission functions $S_n(x)$.

The projective measurements are determined by the laws of propagation and the geometry of the setup. For simplicity, we chose to detect in the positions corresponding to near and far field. This can be done using a lens and placing a detector in the focal plane (far field) and in the plane where the image of the slits is formed (near field). 
In the near field, if the active area of the detector is larger than the image of each slit,  the probability to detect a photon prepared in the state $\ket{\psi}$ as defined above in the position corresponding to $n$th slit image is proportional to $|s_n|^2$. Hence it is easy to see that each of the three positions can be associated with  the measurement operator defined as  
$M_{\text{nf}}(n)=\mu_{\text{nf}}\ket{n}\bra{n},$
where $\mu_{\text{nf}}$ is the normalization factor to be specified later and subscript nf stands for near field. 

The interpretation of measurements in the far field needs more attention. A detection in the position $x$ in the focal plane corresponds to  the projector onto $\ket{k_x}$, which is related to the plane wave propagating in the direction given by the transverse wave vector $k_x=x k/f$. Hence the probability to detect a photon $|\braket{\psi|k_x}|^2$ can be seen as proportional to $\left|\sqrt{\frac{a}{2\pi}}\textrm{sinc}\left(\frac{1}{2}k_x a\right)\braket{\phi(k_x d)|\psi}\right|^2$,  where we introduced $\ket{\phi(\theta)}=\ket{0}+\exp(i \theta)\ket{1}+\exp(i 2 \theta)\ket{2}$. Based on this observation we can define the measurement operator in the far field as: 
$M_{\text{ff}}(\theta) = \mu_{\text{ff}}(\theta)\ket{\phi(\theta)}\bra{\phi(\theta)},$
where $\mu_{\text{ff}}(\theta)$ is the normalization factor,  the phase parameter reads as $\theta=2\pi x d / \lambda f$  and the subscript ff stands for far field. 

The measurement operators $M_{\text{nf}}$ and $M_{\text{ff}}$  can be used to construct rank 7 positive value operator measure (POVM) set allowing one for reconstruction of arbitrary pure state. For this reason we take three near field measurements  $M_\text{nf}(n)$,  $n=0,1,2$ and six far field operators $M_\text{ff}(\theta)$ corresponding to  $\theta=\{0,\pi,2\pi/3,-2\pi/3,5\pi/3,-5\pi/3 \}$. This specific choice requires renormalization, which can be done  when $6\mu_{\text{ff}}(\theta)=\mu_{\text{nf}}=1/2$.


\section{Quantum game}
The ability to encode and measure qutrit states can be utilized  to demonstrate Aharon and Vaidman's quantum game \cite{Aharon2008}. A classical strategy allows Alice for at most $50\%$ chance to win. On the other hand, when she uses quantum particles her chance rises above this limit and ideally reaches $100\%$, when she chooses her initial state to be $\ket{\sa}=\frac{1}{\sqrt{3}}\left(\ket{0}+\ket{1}+\ket{2}\right)$ in the first turn of the game.
In the second turn, assuming Bob has access to slits (boxes) $0$ and $2$, if he decides to check if the photon is passing though slit number $0$, he does  a projective measurement on the state $\ket{0}$. If he finds the particle, then his state becomes $\ket{\sbp}=\ket{0}$, otherwise $\ket{\sbn}=\frac{1}{\sqrt{2}}\left(\ket{1}+\ket{2}\right)$.  The photon detection results in losing the photon from the system, otherwise the photon goes through opened slits. This can be simulated by blocking slit number 0, which will allow us to simulate all those cases when Bob's detector did not click. On the other hand, the case of finding a photon in slit number 0 can be simulated by closing all others.
Next, in the  third turn of the game Alice makes a projective measurement  on $\ket{\psi_{Am}}=\frac{1}{\sqrt{3}}\left(\ket{0}-\ket{1}+\ket{2}\right)$. This can be done by placing the detector D1 in the far field plane in the position corresponding to POVM element $M_\text{ff}(\theta=\pi)$. If Alice detects a particle, she accepts the game trial, and if she does not, she cancels it. 
Now it is clear that Alice cannot lose as whenever Bob does not detect a photon, the state after the second turn is $\ket{\sbn}$ and Alice's detector never clicks as $|\braket{\sbn|\psi_{Am}}|^2=0$. If Bob found the particle in a slit $0$ and tried to leave no trace of that, Alice has $|\braket{\sbp|\psi_{Am}}|^2=1/3$ chance to detect it after that. The same reasoning holds if Bob chooses the slit number $2$. 
\section{Experiment}

The experimental setup is depicted in \figref{fig:experiment}. We used two single photon sources: heralded parametric downconversion (PDC) source based on periodically polled potassium titanyl phosphate (PPKTP) crystal (PDC-SPS) and attenuated HeNe laser (AL-SPS). In order to fulfill the assumption of the plane wave incidence at the slits we used single mode fibers (SMFs) and optics to set the characteristic spatial mode diameter to approximately~ $3$ mm. We control the state of the qutrit using blocking mask (B) to change the configuration of opened slits and tilting slightly the mirror (M) to change the incidence angle $\alpha$.  Under these simplifying assumptions the experimentally possible states are in the following form $\ket{\psi}=\left(\ket{0}+e^{i k d \sin \alpha} \ket{1} + e^{2 i k d \sin \alpha} \ket{2} \right)/\sqrt{3}$. 
The far and near field measurements were implemented by photon counting in the transverse planes at distances of $150$ and $326$ mm, respectively. Large $2$ inch pellicle beamsplitter (BS) was introduced to reduce the disturbance of the setup while changing the planes of detection. Each detector system (D1,D2) was comprised of a multimode fiber mounted on precise motorized stage and Perkin Elemer avalanche photodiode. Step motors were used  to control the transverse position of the fiber with an accuracy of  $1~\mu$m. Counts were registered by the field programmable gate array (FPGA) logic system.

Before simulating the AV game and characterizing prepared states the setup was calibrated. It was done by opening all slits and setting the initial direction of photon propagation to $\alpha=0$, which corresponded to preparing the state $(\ket{0}+\ket{1}+\ket{2})/\sqrt{3}$. Next we measured the photon count rates as a function of the detector position in far- and near-field planes. The results together with the best fits are presented as insets in \figref{fig:experiment}. Blue (online) dots represent experimental data, while the continuous line is a theoretical fit.  The positions corresponding to the far-(near-) field part of POVM are marked with bigger red dots on inset next to detector D1(D2). Note that the smoothed shape of the slits image (near field) is attributed to the finite size of a detector.

\begin{figure}
\includegraphics[width=\columnwidth]{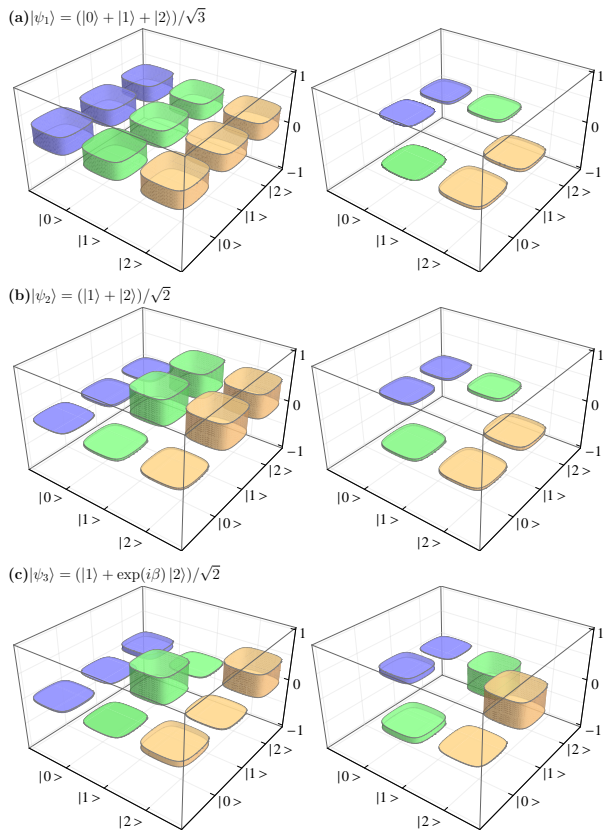}
  \caption{(Color online) States reconstructed based on experimental data shown in Table \ref{tab:counts}. Left (right) column depicts real (imaginary) part of the reconstructed density matrix. The reconstructed phase related to state $\ket{\psi_3}$ was $\beta=1.6$.}
  \label{fig:density}
\end{figure}

\begin{table}
{\begin{tabular}{| c || c | c| c| c|}
\hline
{{Meas.~setting}}/{{Input state}} & ${\ket{\psi_1}}$ & ${\ket{\psi_2}}$ & ${\ket{\psi_3}}$   \\
\hline\hline
${M_\text{ff}(-5\pi/3)}$ 	& 0.097(1) & 0.107(1) &  0.028(1) \\
${M_\text{ff}(-2\pi/3)}$ 	& 0.0017(1) & 0.056(1) &  0.034(1) \\
${M_\text{ff}(0)}$ 		& 0.259(1) & 0.177(1) &  0.176(1) \\
${M_\text{ff}(2\pi/3)}$ 	& 0.0014(1) & 0.040(1) &  0.049(1) \\
${M_\text{ff}(\pi)}$		& 0.031(1) &  0.0010(1) &  0.178 (1)\\
${M_\text{ff}(5\pi/3)}$	& 0.108(2) & 0.117(2) &  0.040(1) \\
${M_\text{nf}(0)}$ 		& 0.167(1) & 0.0027(1) &  0.0030(1) \\
${M_\text{nf}(1)}$ 		& 0.167(1) & 0.260(1) &  0.258(1) \\
${M_\text{nf}(2)}$ 		& 0.165(1) & 0.237(1)&  0.240(1)\\
\hline
\end{tabular}}
\caption{ Photon count probabilities were measured in the positions related to projective measurements $M_{\text{ff}}$ and $M_{\text{nf}}$ for three typical states  $\ket{\psi_1}=\ket{0}+\ket{1}+\ket{2}$, $\ket{\psi_2}=\ket{1}+\ket{2}$, $\ket{\psi_3}=\ket{1}+\ket{2}\exp(1.6 i)$.  
}
\label{tab:counts}
\end{table}
We characterized the prepared qutrit, which has been done resorting to POVM set described earlier and quantum state tomography methods. For this reason the photon counts were measured by placing detectors in positions related to measurements operators $M_{\text{ff}}$ and $M_{\text{nf}}$. Those positions are marked with red dots on inset plots in \figref{fig:experiment}. In order to justify the results of the quantum game we took three typical states: $\ket{\psi_1}=(\ket{0}+\ket{1}+\ket{2})/\sqrt{3}$, $\ket{\psi_2}=(\ket{1}+\ket{2})/\sqrt{2}$, $\ket{\psi_3}=(\ket{1}+\ket{2}\exp(i \beta))/\sqrt{2}$.  For the first state all slits were open, for the second one a slit number 0 was closed and for the last state the propagation direction of the photon  was modified in order to introduce a phase $\beta$. These measurements were done using AL-SPS, its outcomes are gathered in Table \ref{tab:counts} and the results of tomographic reconstruction using the maximal likelihood method are depicted in \figref{fig:density}. 
It is seen in \figref{fig:density}(a,b) that for the states $\ket{\psi_1}$ and $\ket{\psi_2}$ the real part dominates as there is no phase present. On the other hand changing the initial photon direction $\alpha$ it was possible to introduce the phase as is apparent in  \figref{fig:density}(c) as the imaginary bars are significant.  Ideally, the imaginary part of the density matrix for states $\ket{\psi_1}$ and $\ket{\psi_2}$ is zero. Here, the nonzero height is attributed to the noise originating from dark counts, stray light and imperfect positioning. 

\begin{figure}
\includegraphics[width=\columnwidth]{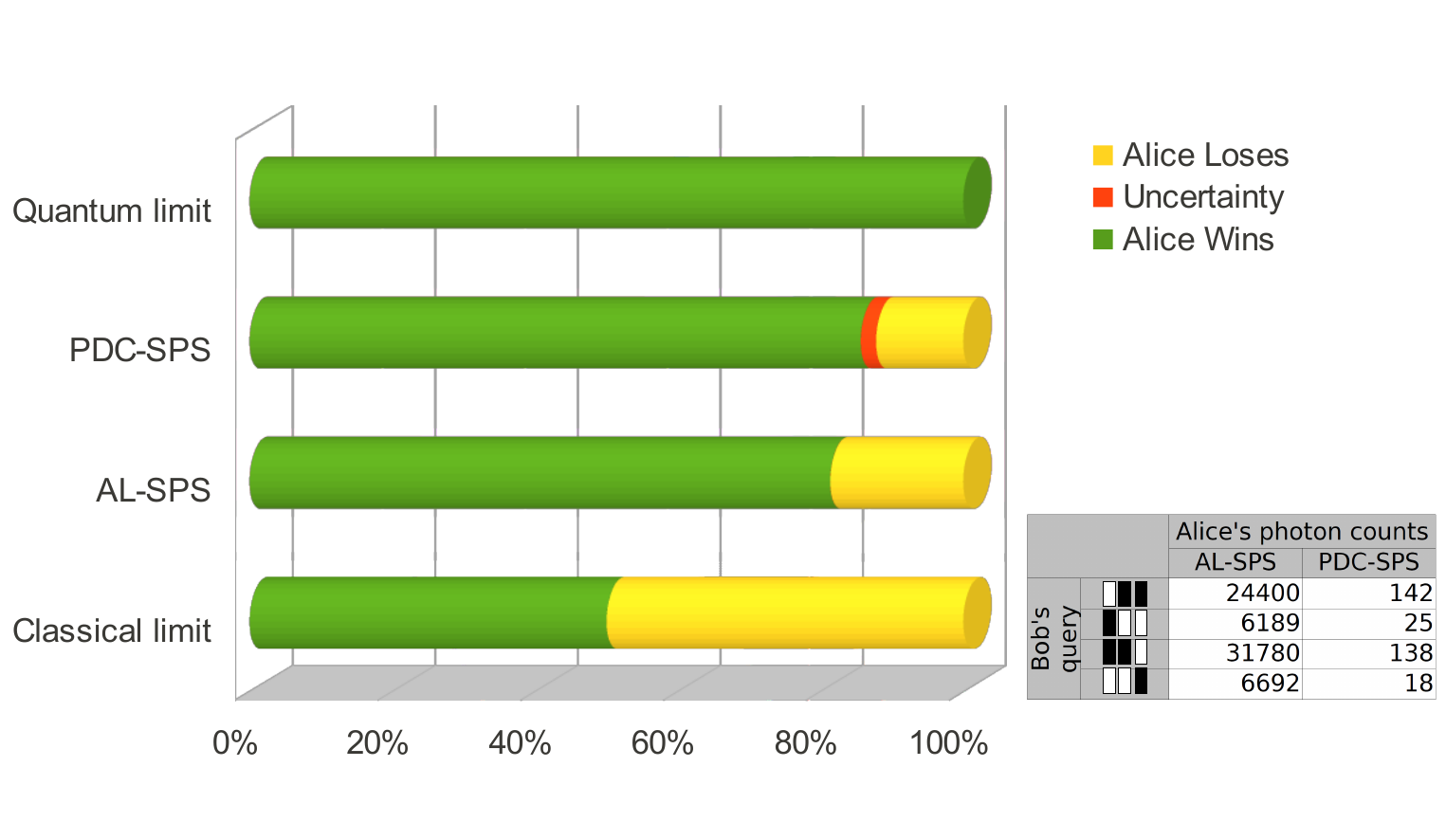}
\caption{ (Color online) Experimental and theoretical best classical and best quantum winning trials in
the quantum game. The simulation shows the quantum advantage over classical limit of 50\% as Alice wins in 82\%  (87\%) of accepted game trials when using AL-SPS (PDC-SPS). Table shows the number of photon counts measured by Alice for each of possible actions by Bob and his measurement outcome. For AL-SPS (PDC-SPS) photons were collected for $2$ s ( $1$ min, coincidence window $1$ ns).}
\label{fig:qg}
\end{figure}

For the quantum game the qutrit was prepared in $\ket{\psi_A}=  \ket{\psi_1}$, which was characterized before, see \figref{fig:density}(a). We simulated all possible scenarios of Bob's measurement using PDC-SPS and AL-SPS. The measured photon counts are presented in table as an inset of \figref{fig:qg}. In the perfect case one expects no counts when two slits are open and Alice sets her detector in the far field plane in the position related to $M_\text{ff}(\theta=\pi)$. This corresponds to the first local maximum marked with red dot on plot related to far field inserted in \figref{fig:experiment}. Here, the measured counts are attributed to the finite size of the multimode fiber core, dark counts and stray light. We estimate that the former two contribute approximately 3 coincidence counts per 2 seconds. Despite the background noise from stray light and dark counts, and the slight non-ideal properties of Alice’s measurements, she won in 87\% of the accepted trials using PDC-SPS and in 82\% of the accepted trials using AL-SPS. Note that the overall efficiency of the game, which is due to experimental deficiencies including the photon collection, the detection efficiencies and number of Alice’s detectors, will only limit the number of accepted trials, but not the percentage of winning trials.

%
%

\section{Discussion}
In conclusion, we experimentally presented a simple way to implement a qutrit system into a single photon's spatial degree of freedom, which allowed us to perform state tomography and simulate the AV quantum game. The encoding part resorted to the Young type experiment, where a photon passes through 3 slits, which defined its state. By controlling an initial propagation direction of a photon  and configuration of the slits it was possible to encode a certain class of states.
Our state reconstruction technique was based on a small number of measurements over a short period of time, which makes our method stable and time efficient in contrast to Ref.~\cite{Taguchi2009},


%


\section{Acknowledgements}
The authors  acknowledge insightful discussions with C.~Fuchs, M.~Graydon and G.~Noel Tabia from Perimeter Institute and funding from NSERC (CGS, QuantumWorks, Discovery, USRA), Ontario Ministry of Research and Innovation (ERA program), CIFAR, Industry Canada and the CFI. PK acknowledges fruitful discussions with R.~Demkowicz-Dobrzanski from Warsaw University, support by the Foundation for Polish Science TEAM project cofinanced by the EU European Regional Development Fund and the Mobility Plus project financed by Polish Ministry of 	Science and Higher Education. AC acknowledges support by MICINN Project No.~FIS2008-05596 and the Wenner-Gren Foundation. 

\end{document}